\documentclass[aip,preprint,prb]{revtex4}
\usepackage{amssymb}
\usepackage{graphicx}

\begin{document}
\title{Magnetic properties of the insulating ferromagnetic phase in strained Pr$_{0.6}$Ca$_{0.4}$MnO$_3$ thin films}
\author{T. Mertelj$^{1,2}$, R. Yusupov$^{1}$, M. Filippi$^{3}$ , W. Prellier$^{3}$ and D. Mihailovic$^{1,2}$}
\author{}
\date{\today}

\begin{abstract}
Bulk magnetization in Pr$_{0.6}$Ca$_{0.4}$MnO$_3$ thin films with tensile (SrTiO$_3$) and compressive (LaAlO$_3$) substrate-induced strain is compared to the magnetooptical Kerr effect (MOKE) measurements. In the absence of an external magnetic field, in both films, a stable ferromagnetic insulating majority phase coexists with an antiferromagnetic insulating phase below $\sim$120K. MOKE  measurements indicate that at 5K a metastable ferromagnetic metallic (FM) phase is formed at the surface of the stretched film in a magnetic field below 1.1T already, while in the bulk the FM phase starts to form in the field above $\sim$ 4T in both films.

\end{abstract}

\affiliation{$^{1}$Jozef Stefan Institute, Jamova 39, 1000
Ljubljana, Slovenia}

\affiliation{$^{2}$Faculty of Mathematics and Physics, Univ. of
Ljubljana, 1000 Ljubljana, Slovenia}

\affiliation{$^{3}$Laboratoire CRISMAT, CNRS UMR 6508, Bd du Marechal Juin, F-14050 Caen Cedex, France}

\maketitle

In thin films of the giant magnetoresistive manganites the substrate-induced strain is an important parameter influencing their electronic and magnetic properties, due to strong interplay of lattice, charge and spin degrees of freedom. In the (Pr,Ca)MnO$_3$ system the strain influences not only the magnetic anisotropy\cite{SuzukiHwang1997}, but more importantly the equilibrium between ferromagnetic metallic (FM) and insulating phases\cite{SaurelBrulet2006} resulting in a spectacular decrease of the critical magnetic field required to destroy the insulating state in Pr$_{0.5}$Ca$_{0.5}$MnO$_3$ thin films.\cite{PrellierHaghiri2000} 

The ground state of bulk Pr$_{1-x}$Ca$_{x}$MnO$_3$ for $x\geq0.3$ is insulating\cite{TomiokaAsamitsu1996}. Small angle neutron scattering results indicate that in the insulating state at low temperatures antiferromagnetic (AF) and ferromagnetic insulating (FI) phases coexist and are converted to the FM phase upon application of a magnetic field.\cite{SaurelBrulet2006} The critical magnetic field required to destroy the insulating phases increases with Ca content and rises from a few T \cite{TomiokaAsamitsu1996} for $x=0.3$ to over $\sim20$T for $x=0.5$\cite{TokunagaMiura1998}.
Surprisingly, this trend is not observed when comparing $x=0.5$ and $x=0.4$ thin films \cite{PrellierSimon2001}. While for $x=0.5$ the critical field for the resistivity transition can be as low as $\sim5$T, no transition has been observed for $x=0.4$ in field up to 9T which is larger than the bulk value\cite{TomiokaAsamitsu1996}. 

Since the resistivity transition is percolative, an absence of the transition does not necessarily indicate the absence of a partial destruction of the insulating state in a moderate magnetic field. Here we report low temperature DC magnetization measurements in Pr$_{0.6}$Ca$_{0.4}$MnO$_3$ thin films to investigate the formation of the nonpercolating FM phase. We also compare the DC magnetization to magnetooptical Kerr effect (MOKE) measurements to investigate how surface effects and strain relaxation influence magnetic order and the formation of the FM phase.

DC magnetization was measured in a SQUID magnetometer in a magnetic field up to 5 T.\footnote{Diamagnetic substrate contribution was not subtracted since it only contributes to temperature independent background linear in magnetic field not exceeding 10\% at the highest field.} MOKE measurements were performed in polar geometry in a 1.1T-electromagnet in an optical liquid-He flow cryostat equipped with CaF$_2$ windows. A linearly polarized probe beam from a pulse Ti:Sapphire laser system with the photon energy, $\hbar \omega_{probe}$,  either 1.55 or 3.1 eV and the diameter 220 $\mu$m was analyzed upon reflection from a film by a Wollaston prism and a pair of balanced silicon PIN photodiodes using standard lockin techniques. The known optical window contribution to the complex Kerr angle was subtracted from data after the measurement.

Thin films of Pr$_{0.6}$Ca$_{0.4}$MnO$_3$ (PCMO60) with the thickness of $\sim$3000 \AA   were grown on (001)-oriented
SrTiO$_3$ (STO) and LaAlO$_3$ (LAO) substrates, where STO induces tensile and LAO compressive strain in the film.\cite{NelsonHill2004} The films  were deposited at 720 $^{\circ}$C by the pulsed laser deposition from a stoichiometric sintered pellet under a 200 mTorr atmosphere of flowing oxygen. After deposition, the films were slowly cooled to room temperature under 300 mbar of oxygen.\cite{PrellierSimon2001} PCMO60/STO had [010]-$Pnma$ axis oriented perpendicular to the film plane\cite{PrellierSimon2001} while PCMO60/LAO was [101] oriented.

Contrary to bulk PCMO60,\cite{TomiokaAsamitsu1996} we find that the thin films are ferromagnetic and insulating below $\sim120$K in the absence of an external magnetic field due to substrate-induced strain effects.\cite{NelsonHill2004} 
The temperature dependence of the magnetization is shown in Fig. 1 and is similar to previous results\cite{NelsonHill2004} with the ferromagnetism appearing below $\sim$120K in both samples. Above $\sim$120K the PCMO60/STO sample shows Curie-Weiss behaviour while the PCMO60/LAO sample shows a small temperature independent ferromagnetic magnetization persisting up to temperatures higher than 270K. A similar ferromagnetic phase has been observed also in $x=0.5$ thin films on LAO with a Curie temperature of $\sim240$K.\cite{HaghiriHervieu2000}

Magnetic hysteresis loops in field up to 5T at temperature 5K are shown in Fig. 1 c) and d). They are characterized by a rather large magnetization during the initial field rise indicating that a majority of the film is ferromagnetically ordered already in a magnetic field of 1-2 T. According  to resistivity measurements in this field range\cite{PrellierSimon2001} this phase is the FI phase\cite{SaurelBrulet2006,HaghiriHervieu2000}, since it is not plausible that the majority phase, which is ferromagnetic, would not percolate. In both samples we observe an onset of formation of an additional ferromagnetic phase at $\sim4$T during the first sweep to 5T. We believe that this phase is the FM phase.  The process is irreversible and magnetic hysteresis loops in a moderate field are permanently altered after first application of the highest field (see Fig. 2). The change is most prominent in the PCMO60/STO sample in the field parallel to the film plane, and much smaller in the PCMO60/LAO sample, where there is almost no difference for different magnetic-field directions.

In Fig. 2 we show magnetization hysteresis loops in 1.1-T maximum field. Loops in both samples reveal remanent magnetization for both orientations of the field. For the parallel field the hysteresis curves taken after zero field cooling (ZFC) are very similar in both samples. While there is almost no anisotropy with respect to the magnetic field direction in the PCMO60/LAO sample, both the remanence and coercitivity are smaller in the PCMO60/STO sample in the perpendicular magnetic field indicating predominantly in-plane magnetization with a small out of plane component, assumed to be due to the Bloch walls.

After the application of 5-T field the remanence increases in both samples for both field directions while the coercitivity significantly increases only in the PCMO60/STO sample for the perpendicular field direction. Anisotropy in the PCMO60/STO sample becomes even more pronounced while the response of the PCMO60/LAO film remains almost isotropic.

Next we compare the 1.1-T magnetization loops, which are bulk sensitive, with MOKE loops, which are surface sensitive. According to optical data\cite{OkimotoTomioka1998} the light penetration depth is $\sim$500\AA\footnote{The value is a rough estimate based on reflectivity from ref. [\cite{OkimotoTomioka1998}]} at 1.55-eV photon energy (PE) in Pr$_{0.6}$Ca$_{0.4}$MnO$_3$. The Kerr rotation, $\phi _\mathrm{K}$, at 1.55-eV PE is linear with magnetic field in both samples (see Fig. 3 a) and c)). Due to a large systematic error, which is also linear in the magnetic field, we conclude that the intrinsic $\phi _\mathrm{K}$ contribution is below our sensitivity at this PE. On the contrary at 3.1-eV PE a nonlinearity of $\phi _\mathrm{K}$ with respect to the magnetic field due to the intrinsic response is clearly observed in both samples. 

In the PCMO60/STO sample the $\phi _\mathrm{K}$ loop is different from the ZFC loop in the perpendicular field, but virtually the same as the magnetization loop after the application of a 5-T field. This is unexpected since the maximum field applied to the sample during MOKE measurements never exceeded 1.1 T. We cannot completely exclude that the FM phase is photoinduced on the surface by the probing light during MOKE measurements. However, since we do not observe anything similar in the PCMO60/LAO sample, we believe that the behaviour is intrinsic and that the FM phase forms in the surface layer already at a magnetic field below 1.1T. This surface layer contribution is not detected by SQUID due to its small volume fraction. Surprisingly, the Kerr ellipticity, $\eta _\mathrm{K}$, shows no (or a much narrower) hysteresis despite the hysteresis in $\phi _\mathrm{K}$ and SQUID magnetization loops. This clearly indicates the presence of two different magnetic phases: one with a large and one without (or very small) hysteresis. We believe that one of these phases corresponds to the FI phase and the other to the FM phase. Both phases are observed also in the SQUID magnetization loop measured after exposure to a 5-T field (Fig. 3 a)) indicating that both are present throughout the whole thickness of the film. Similarly, remanence detected by both Kerr rotation and SQUID magnetization loops indicates that the out-of-plane magnetization component exists not only in the bulk but also near the surface of the film.

In the PCMO60/LAO sample a hysteresis similar to that in the SQUID magnetization loops is not observed in any MOKE loop while the field at which partial saturation occurs is similar in MOKE and SQUID magnetization loops. This strongly suggests that in contrast to the PCMO60/STO sample the magnetization of the top layer in the PCMO60/LAO sample lies completely in the plane of the film. Simultaneously, the remanence of the bulk magnetization in the perpendicular field indicates that at least part of the magnetization away from surface is perpendicular to the film plane. This is consistent with the out-of-plane easy axis observed in very thin Pr$_{0.67}$Sr$_{0.33}$MnO$_3$ films on the LAO substrate\cite{WuRzchowski2000} and measurements in thicker La$_{2/3}$Ca$_{1/3}$MnO$_3$ films\cite{ValenciaBalcells2003} where the magnetization is found to rotate along the film thickness from the out-of-plane direction near the LAO substrate to the in-plane direction near the surface.

There is no evidence from MOKE for the presence of the FM phase in the top layer of the PCMO60/LAO film.

In Fig. 4 we show the difference magnetization ($\Delta M$) obtained by subtracting SQUID magnetization loops measured after and before 5-T field exposure. Surprisingly, in the out-of-plane field the coercitivity and remanence of the loops are very similar in both samples, while there is a large difference when the field is applied in the plane. 

In the PCMO60/LAO sample, $\Delta M$ could be interpreted as the magnetization of the induced FM phase. A narrower more rectangular hysteresis observed in the in-plane field is compatible with an in-plane magnetization direction of the induced-FM phase. This suggests that the FM phase is induced mostly in the top layer of the film. 

The same interpretation of $\Delta M$ is not possible in the PCMO60/STO sample. Here  the hysteresis loop in the parallel field displays regions with the clockwise sense of rotation. Such behavior is possible only if the amount of both FI and FM phases is modified by exposure to the 5-T magnetic field. Despite two contributions to $\Delta M$, we interpret the absence of the saturation of $\Delta M$ in the out-of-plane field as an evidence for predominantly in plane magnetization of the induced FM phase corresponding to the hard-axis magnetic anisotropy which is much larger than in the PCMO60/LAO sample. 

The remanence in the response of the induced FM phase for the out-of-plane field in both samples can be attributed to the Bloch walls between magnetic domains, which have magnetization perpendicular to the film plane, as in the case of the FI phase.

We conclude that in contrast to bulk Pr$_{0.6}$Ca$_{0.4}$MnO$_3$, a large amount of FI phase exists in substrate-strained Pr$_{0.6}$Ca$_{0.4}$MnO$_3$ thin films at low temperatures. This phase shows a change of the magnetic anisotropy from an out-of-plane easy-axis near the substrate to an out-of-plane hard-axis near the surface in the PCMO60/LAO film with the compressive substrate-induced strain. In the PCMO60/STO film, with the tensile substrate-induced strain, the FI phase shows an out-of-plane hard-axis-type anisotropy\footnote{We can not distinguish the  in-plane anisotropy from our data.} across the whole thickness of the film. At low temperatures, a metastable nonpercolating FM phase is induced during the application of an external magnetic field with a bulk threshold field about 4T at 5K. At the surface of the film on the STO substrate MOKE indicates the presence of the FM phase already in a magnetic field below 1.1 T at 5K.

We thank A. Gradi\v{s}ek for help with magnetooptical measurements and V.V. Kabanov for fruitful discussions.  The work was supported within the FP6, project NMP4-CT-2005-517039
(CoMePhS).

\newpage
\section*{Figure captions}

Fig. 1: Temperature dependence of magnetization in a weak magnetic field perpendicular to the film plane (a), (b) and magnetic field dependence of the magnetization at 5K (c), (d).

Fig. 2: Magnetization hysteresis curves in a moderate field after zero field cooling (ZFC) to 5K and after a measurement of a 5-T loop at 5K.

Fig. 3: Comparison of Kerr-rotation and magnetization loops (a), (c) and Kerr-ellipticity and ZFC magnetization loops (b), (d) at 5K. Error bars of $\phi _\mathrm{K}$ at 1.55-eV photon energy represent the systematic error corresponding to a residual linear component, which originates in the uncertainty of the subtracted window Faraday rotation. Error bars are omitted for $\phi _\mathrm{K}$ at 3.1-eV since the residual linear component has the same magnitude as at 1.55-eV. The systematic error of $\eta _\mathrm{K}$ is negligible in comparison to the noise present in the data.

Fig. 4: The difference between magnetization loops measured after and before 5-T field exposure.

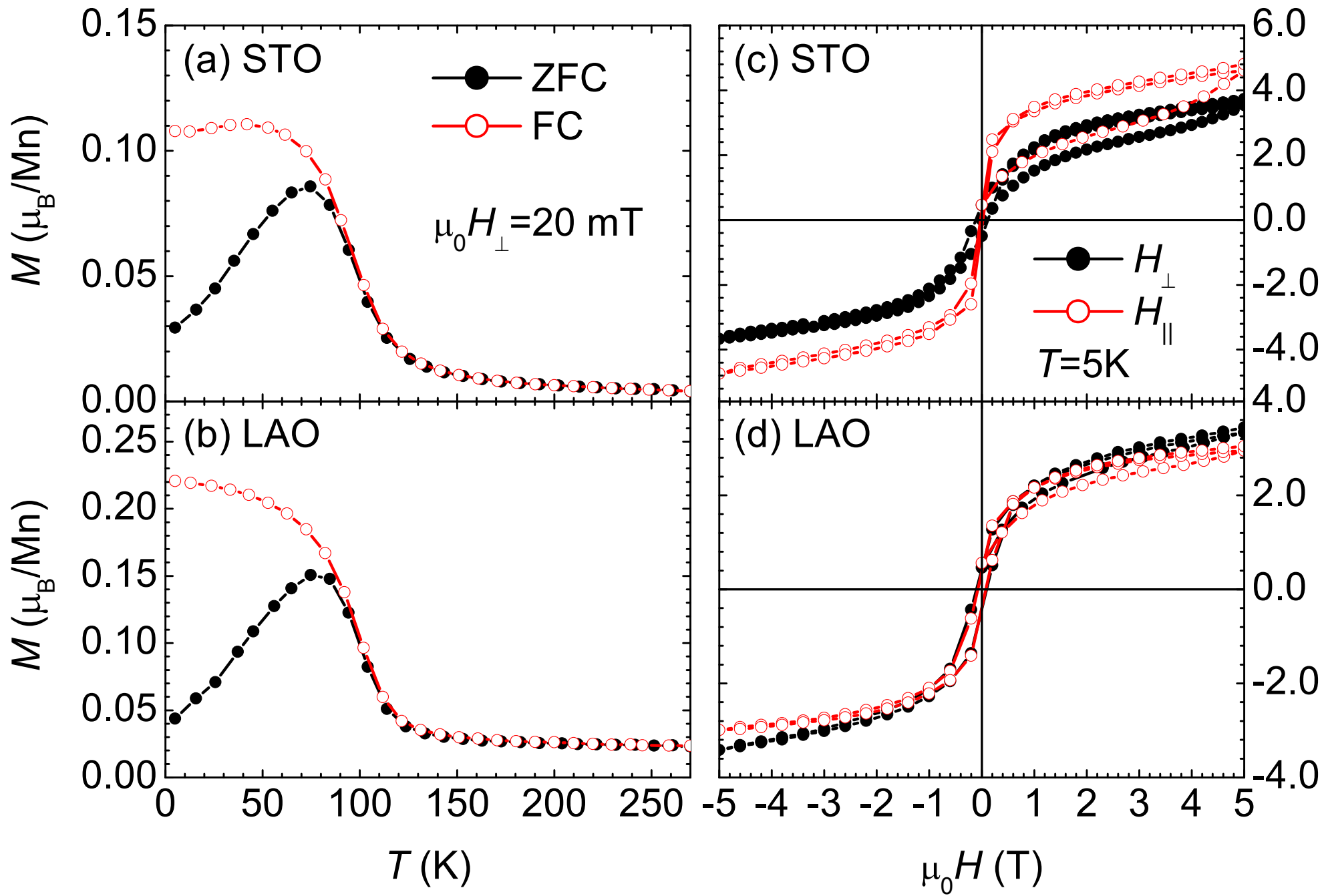
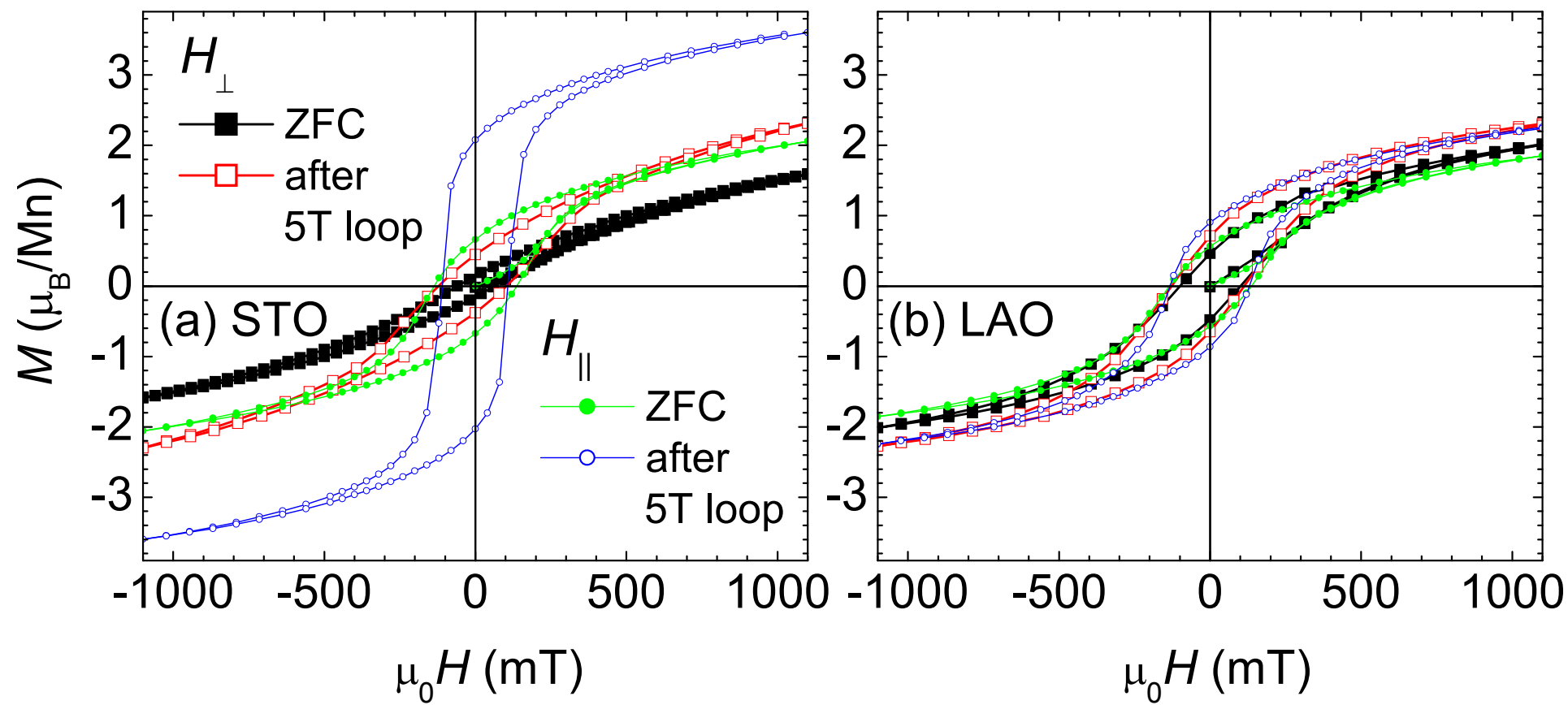
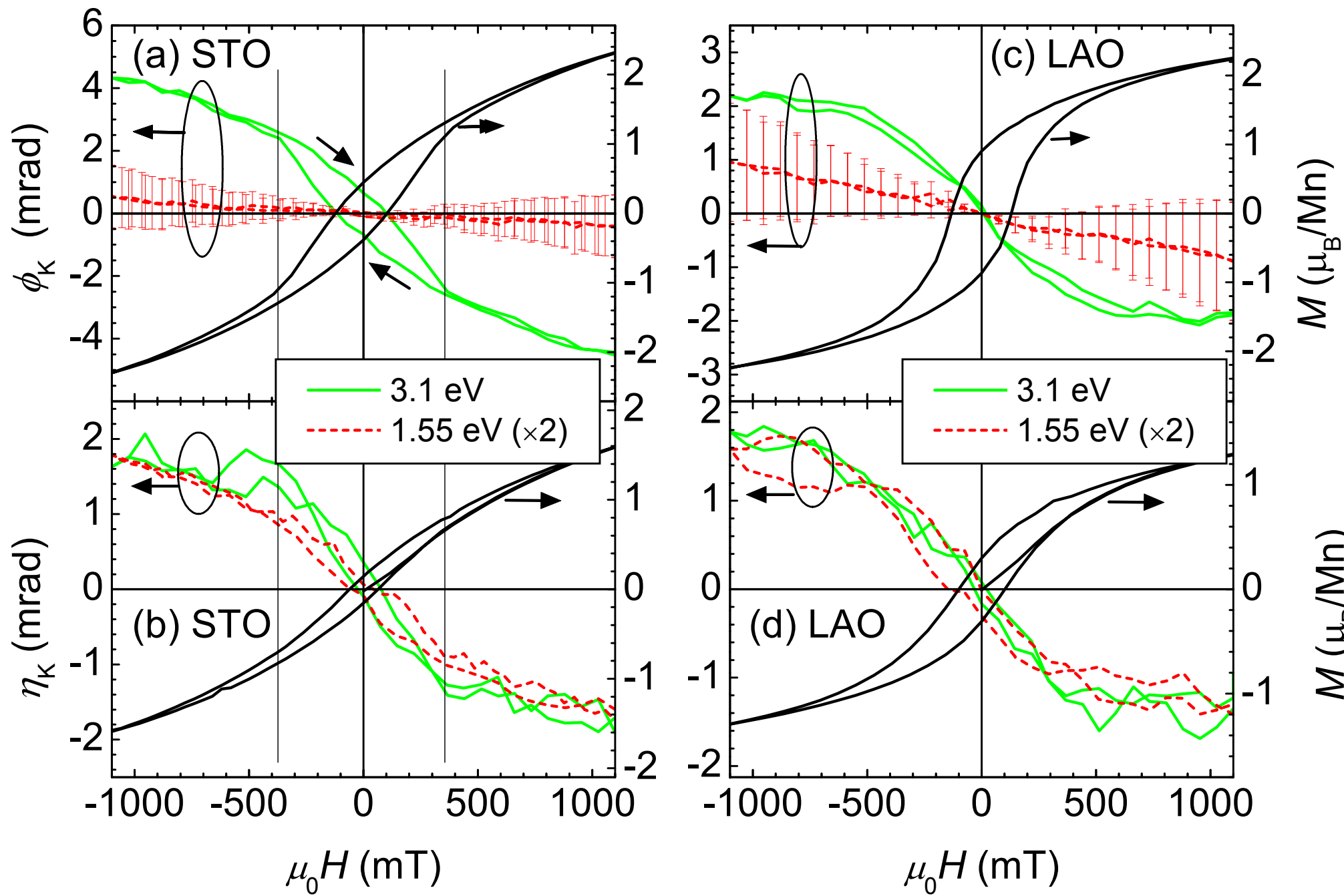
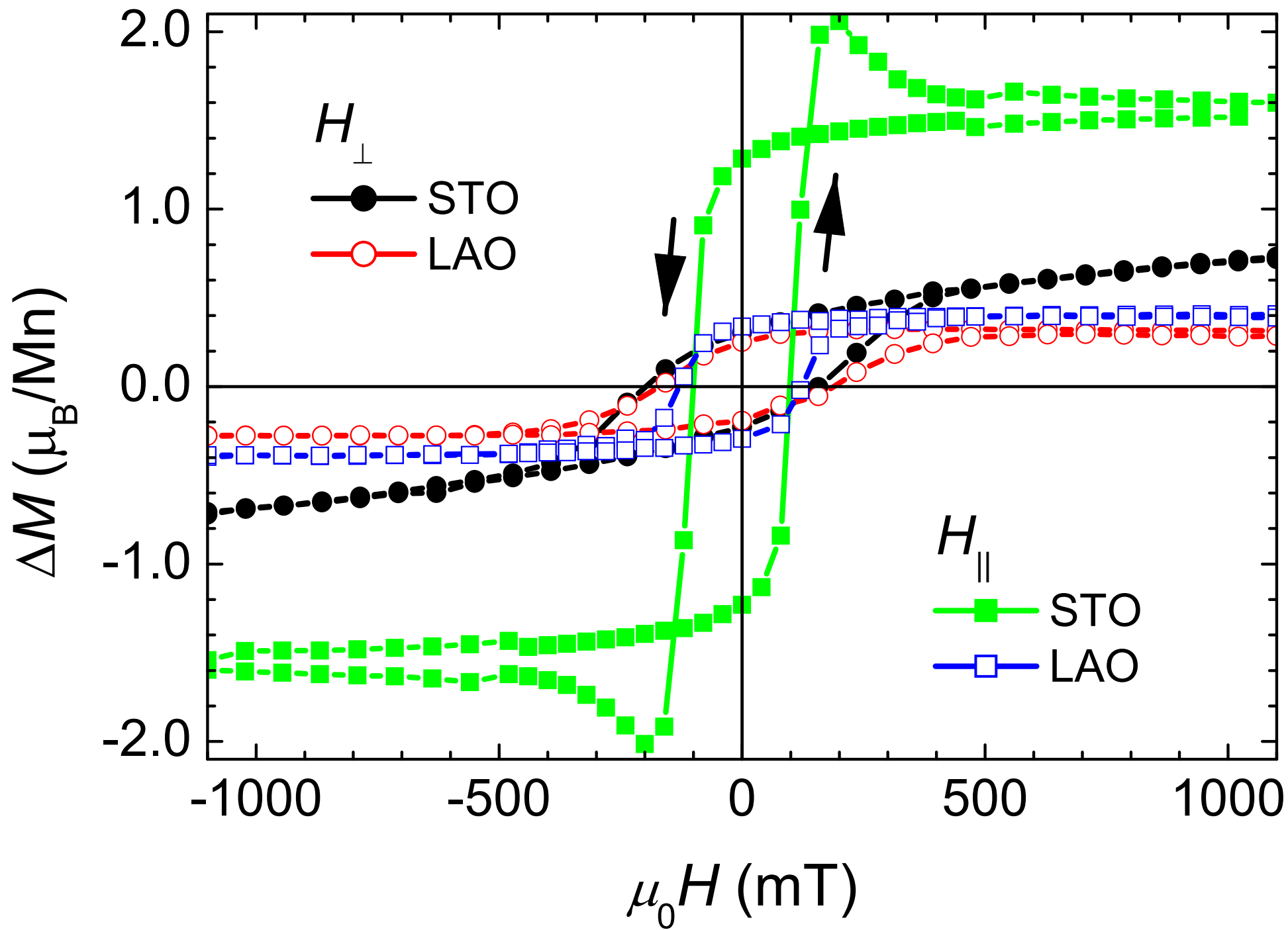
\end{document}